\documentclass[prl,twocolumn,showpacs,amsmath,amssymb]{revtex4-1}

\usepackage{graphicx}
\usepackage{subfigure}
\usepackage{dcolumn}
\usepackage{array}
\usepackage{bm}

\def\ub{\underline}
\def\mcc{\multicolumn{1}{c}}

\usepackage[usenames,dvipsnames]{color}

\begin{document}

\title{\bf{Half-Heusler semiconductors as piezoelectrics} \\[11pt] }

\author{Anindya Roy}
\thanks{These authors contributed equally to this work.}
\affiliation{Department of Physics and Astronomy\\
Rutgers University, Piscataway, NJ 08854}
\author{Joseph W. Bennett}
\thanks{These authors contributed equally to this work.}
\affiliation{Department of Physics and Astronomy\\
Rutgers University, Piscataway, NJ 08854}
\author{Karin M. Rabe}
\affiliation{Department of Physics and Astronomy\\
Rutgers University, Piscataway, NJ 08854}
\author{David Vanderbilt}
\affiliation{Department of Physics and Astronomy\\
Rutgers University, Piscataway, NJ 08854}
\date{\today}

\begin{abstract}
We use a first-principles rational-design approach to demonstrate the
potential of semiconducting half-Heusler compounds as a
previously-unrecognized class of piezoelectric materials. We perform a
high-throughput scan of a large number of compounds, testing for
insulating character and calculating structural, dielectric, and
piezoelectric properties. Our results provide guidance for the
experimental realization and characterization of high-performance
materials in this class that may be suitable for practical
applications.
\end{abstract}
\maketitle

\marginparwidth 2.7in
\marginparsep 0.5in
\def\dvm#1{\marginpar{\small DV: #1}}
\def\ar#1{\marginpar{\small AR: #1}}
\def\jwb#1{\marginpar{\small JWB: #1}}
\def\kmr#1{\marginpar{\small KMR: #1}}
\def\scr{\scriptsize}

One of the central challenges in materials science is the design of
multifunctional materials, in which large responses are produced by
applied fields and stresses. A rapidly developing paradigm for the
rational design of such materials is based on the first-principles
study of a large family of materials.  First principles calculations
of structure and properties are used to explore the microscopic
origins of the functional properties of interest, and this information
is used to guide the computational screening of many compounds, in
both equilibrium and metastable structures, to identify promising
candidate systems. A prototypical example is the optimization of
piezoelectricity in perovskite oxides. The piezoelectric coefficients,
describing the strain induced by an applied electric field or
equivalently a voltage difference induced by applied stress, can be
readily computed from first principles \cite{Wu05p035105,
Baroni01p515}. Screening of a large number of systems has identified
candidate materials for high-performance actuator and sensor
applications
\cite{Fischer06p641,Hautier10p3762}.

\begin{figure}
\includegraphics[width=2.0in]{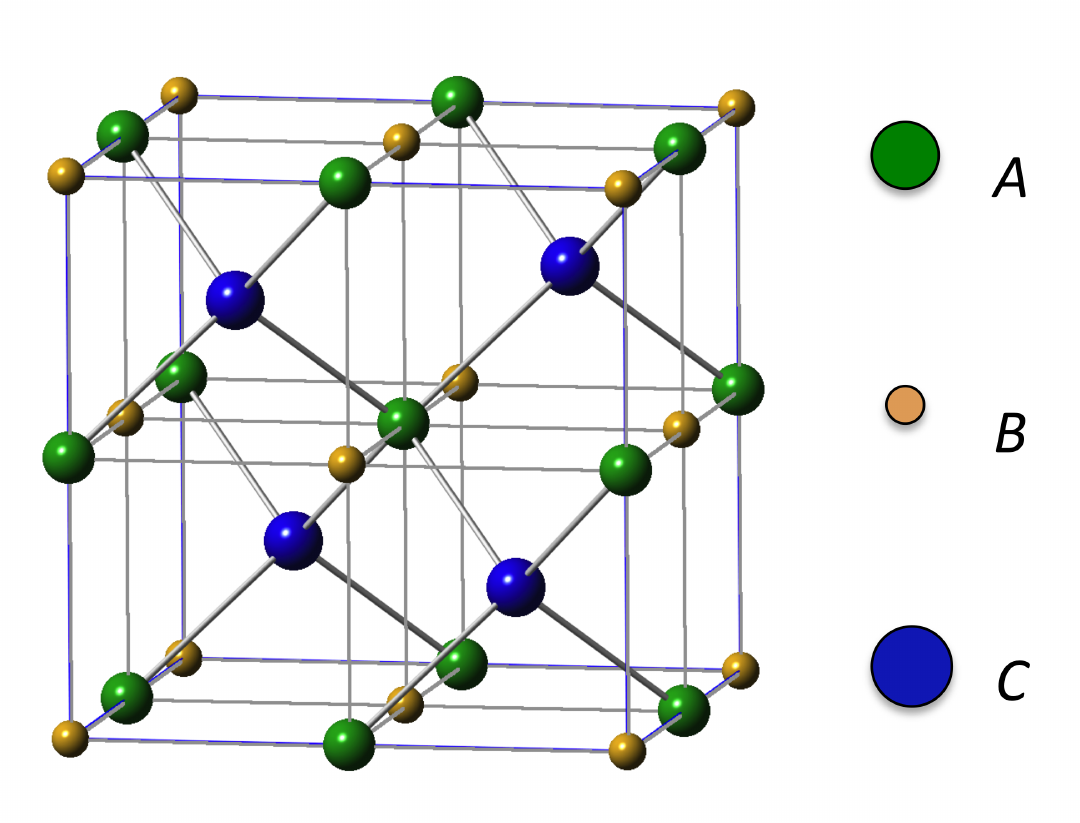}
\caption{{(Color online) The $ABC$ half-Heusler structure: $A$ (green)
    and $B$ (orange) are arranged in a rocksalt lattice, with the
    tetrahedral coordination of $C$ (blue) by $A$ shown.}}
    \label{fig:Figure1}
\end{figure}

Another large family of materials is that of the $ABC$ half-Heusler
compounds (MgAgAs structure type, also called semi-Heusler or
Juza-Nowotny compounds~\cite{Nowotny50p488}), with almost 150 distinct
compounds reported in the {\it Inorganic Crystal Structural Database}
(ICSD)~\cite{Belsky02p364}. The half-Heusler structure, shown in
Fig.~\ref{fig:Figure1}, has $F{\bar4}3m$ symmetry and can be viewed as
a rocksalt lattice formed from two of the three constituent atoms (at
Wyckoff positions 4a and 4b), with the third atom filling half of the
tetrahedral interstitial sites (either Wyckoff position 4c or 4d); it
is related to the $A_2BC$ Heusler structure by removal of one of the
$A$ sublattices, and can alternatively be viewed as a stuffed $AC$ or
$BC$ zincblende structure~\cite{Kandpal06p776}.

Following an initial wave of interest stimulated by the observation of
half-metallic ferromagnetic behavior in half-Heusler
compounds~\cite{deGroot83p2024}, there has been a resurgence of
interest in these compounds as materials that can display topological
properties~\cite{Chadov10p541, Lin10p546} or be tailored for uses as
diverse as components in spintronic devices~\cite{Felser07p668} and
high-performance thermoelectrics~\cite{Shen01p4165, Nolas06p199,
Balke11p702}. The semiconducting half-Heusler compounds are of
particular interest \cite{Ogut95p10443,Wood85p2570}.  As insulators,
these can exhibit functional properties associated with electric
polarization, but these properties have received very little
attention. In fact, while the $F\bar{4}3m$ space group of the
half-Heusler structure allows a nonzero piezoelectric response, no
measurements of piezoelectricity in these systems have yet been
reported in the literature.

In this paper, we use first-principles methods to predict the
piezoelectric response and related properties of half-Heusler
compounds. We present these predictions first for compounds already
reported in the half-Heusler structure, and then perform a
high-throughput analysis of a much larger set of candidate
combinations, identifying high-performance compounds for practical
application. We find that half-Heusler compounds exhibit a wide range
of shear piezoelectric constants; the highest values (found for as-yet
hypothetical compounds) are well above $d_{14}$ = 200\,pC/N. This
compares well to known piezoelectric oxides such as PZT with $d_{33}
\approx 300\,$pC/N and ZnO with $d_{33} \approx 10\,$pC/N.  In
addition, the diversity of possible combinations could allow for other
desirable functional properties to couple to piezoelectricity. Through
targeted synthesis, which might include compositional substitution,
epitaxial growth or artificial structuring, half-Heusler compounds
could thus be developed as a valuable class of piezoelectric
materials, much like the perovskite oxides.

First-principles calculations are performed with the ABINIT
package~\cite{Gonze02p478,Gonze05p558,Gonze09p2582} using the local
density approximation (LDA) and an 8$\times$8$\times$8 Monkhorst-Pack
sampling of the fcc Brillouin
zone~\cite{Monkhorst76p5188}. Optimized~\cite{Rappe90p1227} designed
non-local~\cite{Ramer99p12471} norm-conserving pseudopotentials,
generated using the OPIUM code~\cite{opium}, are employed, with a
plane-wave cutoff of 25\,Ha.

For the high-throughput search, we consider combinations of three
distinct elements $ABC$. We limit the search to combinations with a
total of 8 $s$ and $p$ valence electrons, since we expect this to
improve the likelihood of band gap formation~\cite{Kandpal06p776}.
We also require that there be at least one and at most two $p$-block
elements among the three constituents, with the remainder coming
from the $s$ and/or $d$ blocks of the periodic table. We use Roman
numerals to denote the valences of the constituent elements
(including $d$ electrons) to classify the combinations into families
with 8 valence electrons (I-I-VI, I-II-V, I-III-IV, II-II-IV, and
II-III-III), 18 valence electrons (XI-I-VI, XI-II-V, XI-III-IV,
I-XII-V, II-XII-IV, III-XII-III, X-II-VI, X-III-V, and X-IV-IV), or
28 valence electrons (X-XII-VI, XI-XI-VI, XI-XII-V and XII-XII-IV).
Among these families, we consider members constructed from the
selections I=(Li, Na, K), II=(Be, Mg, Ca, Sr, Ba), III=(B, Al, Ga,
In, Sc, Y), IV=(C, Si, Ge, Sn, Pb, Ti, Zr, Hf), V=(N, P, As, Sb, Bi),
VI=(O, S, Se, Te), X=(Ni, Pd, Pt), XI=(Cu, Ag, Au), and XII=(Zn,
Cd).  This generates a total of 987 candidate combinations to be
searched.

We first consider the 38 combinations in our search set that have been
experimentally reported in the ICSD in the half-Heusler structure. For
each combination, we optimize the lattice constant for each of the
three structural variants $\ub{A}BC$, $A\ub{B}C$, and $AB\ub{C}$,
where the underscore indicates the unique element that is
tetrahedrally coordinated by the other two elements. First-principles
results show that 27 of them, listed in Table~\ref{table:Table1}, are
insulating.  The predicted lowest-energy structural variant is
indicated in the first column.  We find that for the five compounds
(LiZnAs, AuScSn, NiScSb, PdScSb and PtYSb) for which a refined
structure, including R value and temperature factors, is available in
ICSD, our prediction agrees with the experimentally observed variant.
Most of the computed equilibrium lattice constants given in
Table~\ref{table:Table1} are in excellent agreement with
experiment. There are a few exceptions, LiInSi, LiGaSi and LiZnP, for
which the DFT values, confirmed by independent all-electron
calculations, differ significantly from the experimentally measured
values; further experimental investigation of these cases is
warranted. The computed gaps range from 0.07\,eV for AuYPb and LiInSi
to 1.55\,eV for LiMgAs; measured values are expected to be higher
given the well-known tendency of DFT to underestimate band gaps.

For each compound, we perform a linear-response calculation using
density-functional perturbation theory (DFPT)~\cite{Gonze97p10355}, as
implemented in ABINIT, to compute the electronic dielectric constant
$\epsilon_\infty$, dynamical charges and zone-center phonon
frequencies and eigenvectors, from which we obtain the zero-stress
static dielectric constant $\epsilon_{0}$~\cite{born54p268}, reported
in Table~\ref{table:Table1}.  Calculations of the strain response
\cite{Wu05p035105,Hamann05p035117} yield the $C_{44}$ elastic constant
and $e_{14}$ piezoelectric coefficient, also reported in
Table~\ref{table:Table1}.  These results determine
$d_{14}=e_{14}/C_{44}$ and the electromechanical coupling coefficient
$k_{14}$, the conventional figure of merit for piezoelectric
performance, as $k_{14} = |e_{14}| /\sqrt{C_{44}\epsilon_{\rm
fs}\epsilon_0}$, where $\epsilon_{\rm fs}$ is the permittivity of free
space \cite{Wu05p035105}.  We note that $C_{44}$, especially when it
is very small, is difficult to calculate precisely, the more so as it
is also sensitive to the value of the lattice parameter. There is a
corresponding uncertainty in $d_{14}$. However, $k_{14}$ is less
affected as it depends on $C_{44}$ only as 1/$\sqrt{C_{44}}$.  Also,
we note that $\epsilon_\infty$ and therefore $\epsilon_0$ is generally
overestimated in DFT, and $k_{14}\propto \epsilon_0^{-1/2}$, our
calculated values of $k_{14}$ are likely to be underestimates of the
electromechanical coupling coefficients that could be achieved
experimentally.

The computed values of $\epsilon_{0}$, $C_{44}$, $e_{14}$, $d_{14}$
and $k_{14}$ are reported in Table~\ref{table:Table1}.  The values of
the piezoelectric constant $e_{14}$ range from 0.01 to
0.81\,C/m$^2$. Seventeen compounds have $e_{14}>0.16$\,C/m$^2$, the
experimentally measured piezoelectric coefficient of GaAs. The highest
predicted values are for NiTiSn, PtTiSn and LiGaSi. These three
compounds also have the highest values of electromechanical coupling
$k_{14}$, arising from their large $e_{14}$.  LiMgP, LiZnP, and LiZnAs
have similarly high $k_{14}$ values, with lower values of $e_{14}$
compensated by lower values of $\epsilon_0$.

\begin{table}
\begin{center}
\begin{ruledtabular}
\begin{tabular}{l c c c c c c c r}
$ABC$ & \mcc{$a_{\rm expt.}$} & \mcc{$a$}
 & \mcc{$E_{\rm gap}$} & \mcc{$e_{14}$}& \mcc{$d_{14}$}
 & \mcc{$C_{44}$} & \mcc{$k_{14}$} & \mcc{$\epsilon_0$} \\
\hline
LiMg\ub{P} & 6.02 & 5.95 & 1.51 & 0.32 & 7.6 & 0.42 & 0.15 & 12 \\
LiMg\ub{As} & 6.19 & 6.17 & 1.55 & 0.26 & 6.9 & 0.37 & 0.13 & 12 \\
LiMg\ub{Bi} & 6.74 & 6.71 & 0.62 & 0.07 & 2.8 & 0.26 & 0.04 & 16 \\
LiZn\ub{N} & 4.87 & 4.77 & 1.21 & 0.11 & 1.3 & 0.84 & 0.03 & 13 \\
LiZn\ub{P} & 5.78 & 5.59 & 1.15 & 0.44 & 6.6 & 0.67 & 0.15 & 14 \\
LiZn\ub{As} & 5.94 & 5.87 & 1.12 & 0.43 & 8.1 & 0.54 & 0.16 & 15 \\
LiCd\ub{P} & 6.09 & 5.95 & 0.76 & 0.14 & 3.2 & 0.46 & 0.05 & 17 \\
\ub{Au}ScSn & 6.42 & 6.39 & 0.19 & 0.17 & 2.7 & 0.62 & 0.05 & 20 \\
\ub{Au}YPb  & 6.73 & 6.65 & 0.07 & 0.44 & 9.1 & 0.48 & 0.13 & 27 \\
LiGa\ub{Si} & 6.09 & 5.81 & 0.08 & 0.56 & 8.6 & 0.65 & 0.15 & 22 \\
LiIn\ub{Si} & 6.68 & 6.18 & 0.07 & 0.08 & 1.6 & 0.51 & 0.03 & 23 \\
\ub{Ni}TiSn & 5.94 & 5.89 & 0.44 & 0.81 & 12.9 & 0.62 & 0.19 & 30 \\
\ub{Ni}ZrSn & 6.11 & 6.08 & 0.52 & 0.19 & 2.8 & 0.70 & 0.05 & 26 \\
\ub{Ni}HfSn & 6.08 & 6.00 & 0.41 & 0.28 & 3.4 & 0.80 & 0.07 & 24 \\
\ub{Pd}ZrSn & 6.32 & 6.27 & 0.49 & 0.01 & 0.1 & 0.68 & 0.01 & 24 \\
\ub{Pd}HfSn & 6.30 & 6.20 & 0.38 & 0.06 & 0.7 & 0.75 & 0.01 & 22 \\
\ub{Pt}TiSn & 6.16 & 6.15 & 0.82 & 0.64 & 9.5 & 0.67 & 0.16 & 25 \\
\ub{Pt}ZrSn & 6.32 & 6.32 & 1.06 & 0.19 & 2.5 & 0.77 & 0.05 & 20 \\
\ub{Pt}HfSn & 6.31 & 6.25 & 0.99 & 0.23 & 2.7 & 0.83 & 0.06 & 18 \\
\ub{Ni}ScSb & 6.06 & 6.06 & 0.29 & 0.15 & 2.3 & 0.66 & 0.04 & 21 \\
\ub{Ni}ScBi & 6.19 & 6.15 & 0.22 & 0.20 & 3.6 & 0.57 & 0.06 & 24 \\
\ub{Ni}YSb  & 6.31 & 6.28 & 0.34 & 0.20 & 3.6 & 0.55 & 0.06 & 20 \\
\ub{Ni}YBi  & 6.41 & 6.36 & 0.26 & 0.14 & 2.9 & 0.49 & 0.05 & 22 \\
\ub{Pd}ScSb & 6.31 & 6.27 & 0.28 & 0.03 & 0.4 & 0.59 & 0.01 & 20 \\
\ub{Pd}YSb  & 6.53 & 6.47 & 0.26 & 0.23 & 4.2 & 0.53 & 0.08 & 18 \\
\ub{Pt}ScSb & 6.31 & 6.31 & 0.71 & 0.07 & 1.0 & 0.66 & 0.02 & 19 \\
\ub{Pt}YSb  & 6.54 & 6.51 & 0.58 & 0.19 & 3.3 & 0.58 & 0.06 & 19 \\
\end{tabular}
\end{ruledtabular}
\caption{Properties of experimentally synthesized insulating
  half-Heuslers, grouped by family. Experimental lattice constants
  (in \AA) are from the ICSD. Also presented are the
  theoretical lattice constant $a$, band gap $E_{\rm gap}$ (eV),
  piezoelectric coefficients $e_{14}$ (C/m$^2$) and $d_{14}$ (pC/N),
  elastic constant $C_{44}$ (10$^{11}$Pa), electromechanical coupling
  coefficient $k_{14}$, and zero-stress static dielectric constant
  $\epsilon_0$.}
\label{table:Table1}
\end{center}
\end{table}

It is remarkable that no piezoelectric response data for any
half-Heusler compound has yet been reported. With a single-crystal
sample of sufficiently low conductivity, the piezoelectric coefficient
should be readily measurable for most if not all of these compounds.
Moreover, measurements of the dielectric response and elastic
coefficients, which also have not been reported to date, would provide
an additional test of these theoretical predictions and a more
complete characterization of the polarization-related properties of
these otherwise much-studied compounds.

Next, we consider the properties of the full set of 987 hypothetical
and real $ABC$ combinations identified earlier for study. As before,
for each combination we optimize the lattice constant for each of the
variants $\ub{A}BC$, $A\ub{B}C$, and $AB\ub{C}$. Choosing the variant
having the lowest total energy, we determine whether our LDA
calculations predict it to be insulating. Of the 987 combinations, we
find 371 insulators having either 8 or 18 valence electrons, while the
compounds containing 28 valence electrons are all found to be
metallic.  

For the insulators, we perform linear-response calculations using
ABINIT as described above, except that we also compute the $C_{11}$
and $C_{12}$ elastic constants.  This allows us to screen for local
elastic stability by requiring that $C_{11}+ 2C_{12} > 0$, $C_{44} >
0$ and $C_{11} - C_{12} > 0$~\cite{Karki97p8579}. Furthermore, we
calculate phonon frequencies at three additional high-symmetry points
($X, L$ and $W$) and eliminate combinations which exhibit any unstable
modes.  This reduces the combinations further from 371 down to 312.

\begin{figure}
\includegraphics[width=3.0in]{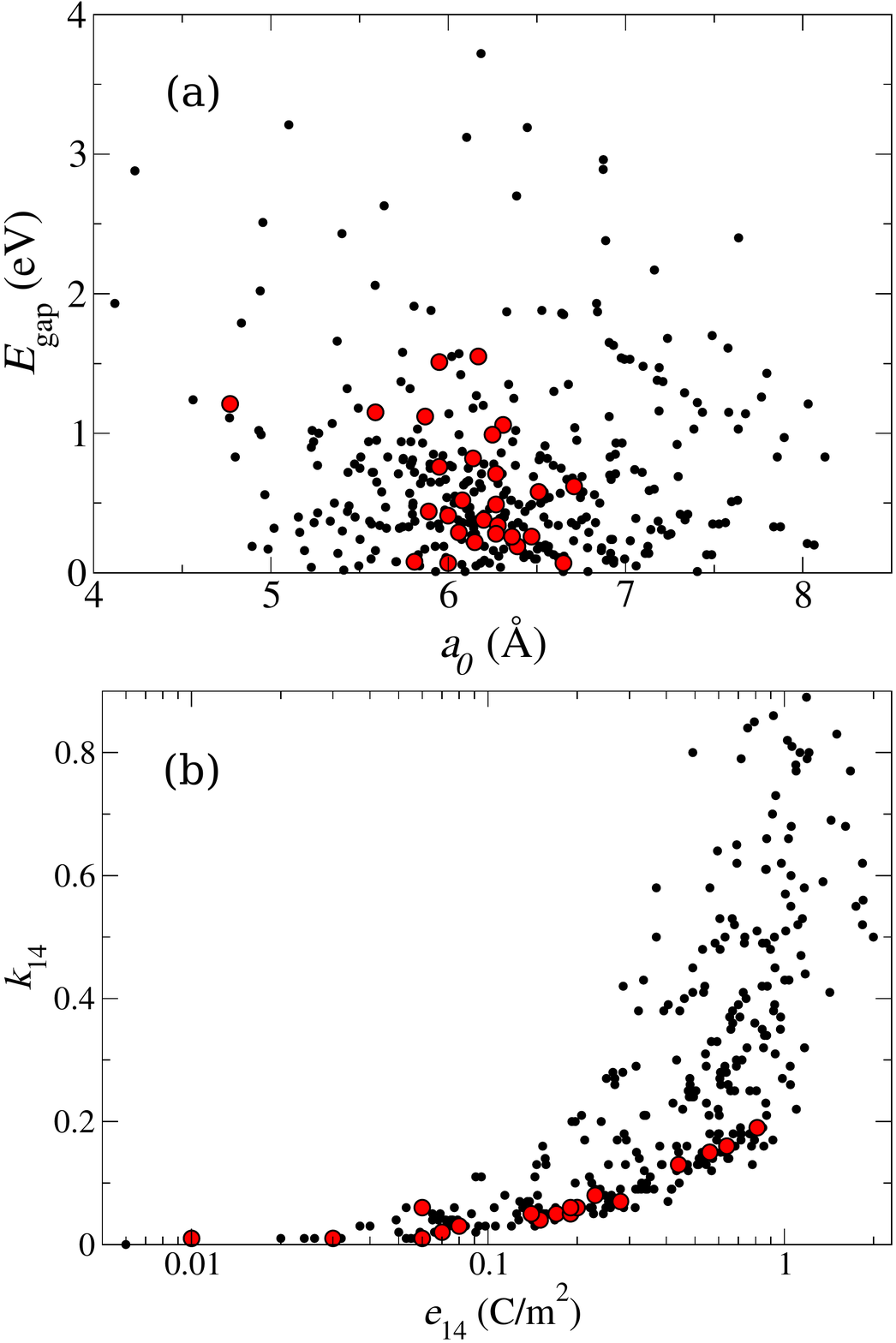}
\caption{(Color online) a)  Cell parameters  (in \AA)  of insulating
    $ABC$ combinations as well as their range of band gaps (in eV) are
    depicted as open black circles. b) Electromechanical  coupling
    factor $k_{14}$ of insulating  $ABC$ combinations as a function of
    piezoelectric  constant $e_{14}$. Known combinations, from
    Table~\ref{table:Table1}, are highlighted as filled red circles. }
\label{fig:Figure2}
\end{figure}

We thus arrive at 312 combinations that are predicted to be insulating
and locally stable in the lowest-energy variant of the three possible
$ABC$ half-Heusler structures. Since DFT tends to underestimate band
gaps, we expect that the actual fraction of insulating structures will
be slightly higher than our calculations would indicate.  The computed
band gaps and lattice parameters for these 312 compounds are shown in
Fig.~\ref{fig:Figure2}(a). We expected no particular correlation
between lattice parameter and band gap, and indeed we find none.  Both
quantities are rather broadly distributed, suggesting that there could
be considerable flexibility in choosing materials over a substantial
range of desired gap or lattice constant.

For these 312 compounds, we compute $\epsilon_{0}$, $C_{44}$, $e_{14}$, $d_{14}$
and $k_{14}$, using the same methods as before.  To give a sense of
how the range of properties in the full set of known and hypothetical
compounds compares with that of the subset of known compounds, we
present a scatter plot of $k_{14}$ vs log $e_{14}$ in
Fig.~\ref{fig:Figure2}(b). It can be seen that there are hypothetical
compounds with $k_{14}$ values well above those of known compounds
(some have $k_{14}$ close to one). The twenty-seven compounds with the
highest values of $k_{14}$ are listed in Table~\ref{table:Table2}.

\begin{table}
\begin{center}
\begin{ruledtabular}
\begin{tabular}{l c c c c c c r}
$ABC$ & \mcc{$a$} & \mcc{$E_{\rm gap}$} & \mcc{$e_{14}$} & \mcc{$d_{14}$} &
\mcc{$C_{44}$} & \mcc{$k_{14}$} & \mcc{$\epsilon_0$} \\
\hline
 {\bf KMg\ub{P}}   & 6.71 & 1.04 & 1.19 & 346.28 & 0.03 & 0.89 & 13\\
 {\bf LiNa\ub{S}}  & 6.10 & 3.12 & 0.92 & 208.22 & 0.04 & 0.86 & 8\\
 LiSr\ub{As}  & 6.84 & 1.93 & 0.79 & 286.27 & 0.03 & 0.85 & 10\\
 {\bf LiNa\ub{Se}} & 6.39 & 2.70 & 0.75 & 227.09 & 0.03 & 0.84 & 8\\
 {\bf NaMg\ub{N}}  & 5.43 & 0.72 & 1.50 & 165.23 & 0.09 & 0.83 & 13\\
 Na\ub{Be}P   & 5.59 & 2.06 & 1.02 & 192.98 & 0.05 & 0.82 & 11\\
 KMg\ub{As}   & 6.92 & 0.67 & 1.06 & 205.75 & 0.05 & 0.81 & 13\\
 KSc\ub{Pb}   & 7.35 & 0.42 & 1.21 & 330.83 & 0.04 & 0.80 & 25\\
 MgBa\ub{Si}  & 7.18 & 0.37 & 1.13 & 290.51 & 0.04 & 0.80 & 21\\
 \ub{Be}BaSn  & 6.86 & 0.16 & 0.49 & 1036.54 & 0.01 & 0.80 & 33\\
 KSc\ub{Sn}   & 7.31 & 0.42 & 1.19 & 294.69 & 0.04 & 0.79 & 24\\
 K\ub{Be}Sb   & 6.51 & 0.81 & 0.72 & 266.50 & 0.03 & 0.79 & 13\\
 MgBa\ub{Ge}  & 7.21 & 0.06 & 1.10 & 266.49 & 0.04 & 0.78 & 21\\
 AgSr\ub{As}  & 6.72 & 0.34 & 1.10 & 163.79 & 0.07 & 0.77 & 14\\
 NaZn\ub{P}   & 6.02 & 0.76 & 1.67 & 163.79 & 0.14 & 0.77 & 16\\
{\bf Na\ub{B}Si}   & 5.43 & 0.44 & 0.94 & 121.79 & 0.06 & 0.73 & 13\\
{\bf KBa\ub{N}}    & 6.48 & 0.69 & 0.91 & 174.57 & 0.05 & 0.70 & 18\\
NaZn\ub{As}   & 6.26 & 0.16 & 1.44 & 98.70  & 0.15 & 0.69 & 18\\
 KY\ub{Ge}    & 7.13 & 0.16 & 1.06 & 178.85 & 0.06 & 0.68 & 25\\
 Ca\ub{B}Al   & 5.81 & 0.37 & 1.61 & 99.88  & 0.16 & 0.68 & 21\\
 KMg\ub{Sb}   & 7.30 & 0.69 & 0.87 & 108.05 & 0.08 & 0.66 & 14\\
 KY\ub{Si}    & 7.11 & 0.14 & 1.03 & 167.49 & 0.06 & 0.66 & 25\\
 \ub{Cu}SrAs  & 6.39 & 0.36 & 0.69 & 179.58 & 0.04 & 0.65 & 19\\
 NaK\ub{O}    & 5.90 & 1.88 & 0.60 & 80.78  & 0.07 & 0.64 & 8\\
 K\ub{Be}Bi   & 6.62 & 0.34 & 0.69 & 127.24 & 0.05 & 0.62 & 16\\
 {\bf KAl\ub{Si}}  & 6.57 & 0.54 & 0.98 & 101.96 & 0.10 & 0.62 & 18\\
ZnY\ub{B}     & 5.78 & 0.33 & 1.84 & 95.35  & 0.19 & 0.62 & 32\\
\end{tabular}
\end{ruledtabular}
\end{center}
\caption{Top half-Heusler candidate compounds, ranked
  according to predicted electromechanical coupling coefficient
  $k_{14}$. Those that remain after applying the filtering criteria,
  as described in the text, are in boldface. Units as
  in Table 1.}
\label{table:Table2}
\end{table}

As a guide for experimental investigation of piezoelectric
half-Heusler compounds, we highlight a selection of candidate
compounds chosen according to relevant practical considerations. We
filter the list to eliminate compounds with toxic (Pb, Cd, As) or
expensive (Be, Pd, Pt, Au, In) elements. In addition, we require a DFT
band gap above 0.4\,eV to favor low sample conductivity. Lastly, to
promote cation ordering into the lowest-energy variant, we require
$\Delta E>0.15$\,eV, where $\Delta E$ is the difference in energy
between the lowest-energy and next-lowest-energy variants. Those
entries in Table II that satisfy these criteria are shown in boldface. Of
the 100 compounds that satisfy these criteria, 79 have
$e_{14}>0.16$\,C/m$^2$, the experimentally measured value of GaAs.

\begin{table}
\begin{center}
\begin{ruledtabular}
\begin{tabular}{l d d d d d d r}
$ABC$ & \mcc{$a$} & \mcc{$E_{\rm gap}$} & \mcc{$e_{14}$} & \mcc{$d_{14}$} & \mcc{$C_{44}$} & \mcc{$k_{14}$} & \mcc{$\epsilon_0$} \\
\hline
Li\ub{Ag}Te & 6.38 & 1.02 & 0.80 & 28.60 & 0.28 & 0.36 & 17\\
Li\ub{Cu}Te & 6.06 & 1.57 & 0.38 &  8.14 & 0.38 & 0.16 & 14\\
LiSc\ub{C}  & 5.23 & 0.90 & 0.34 &  7.73 & 0.44 & 0.12 & 19\\
LiCu\ub{S}  & 5.43 & 1.32 & 0.22 &  4.04 & 0.55 & 0.10 & 10\\
LiMg\ub{Sb} & 6.59 & 1.30 & 0.15 &  4.94 & 0.30 & 0.08 & 14\\
LiIn\ub{Si} & 6.18 & 0.07 & 0.08 &  1.64 & 0.51 & 0.03 & 23\\
\end{tabular}
\end{ruledtabular}
\caption{Piezoelectric properties of the previously uncharacterized
combinations from Ref.~\cite{Zhang12p1425} in which the ground state
is predicted to be the half-Heusler structure. Units as
in Table I.}
\label{table:Table3}
\end{center}
\end{table}

Further investigation of the hypothetical half-Heusler piezoelectrics
hinges on the possibility of experimentally realizing the desired
compounds in the half-Heusler structure. Additional information about
bulk equilibrium $ABC$ phases can be obtained from the ICSD. In the
cases where no phase is reported, it could be that either no stable
bulk phase exists with that stoichiometry, or simply that the relevant
composition has not been studied. $ABC$ combinations are also reported
in several structures other than the half-Heusler structure.
Specifically, we find that twelve combinations listed in Table \ref{table:Table2} are
reported in ICSD with other structures. Nine (KMgP, LiNaS, MgBaSi,
MgBeGe, NaZnP, NaZnAs, KMgSb, NaKO, and KMgAs) are reported in the
$P4/nmm$ PbClF (also called Cu$_2$Sb) structure, one (LiNaSe) in the
$Pnma$ MgSrSi structure, and two (AgSrAs and CuSrAs) in the
$P6_{3}/mmc$ ZrBeSi structure. One (NaZnAs) is reported in both
$P4/nmmm$ and $Fm{\bar 3}m$ structures.


First-principles calculations of the total energy of alternative
structures can also be used both to predict ground state structures
and to identify systems with piezoelectric structures as low-energy
alternatives that would be suitable candidates for stabilization
through compositional substitution or epitaxial strain. In a recent
first-principles study \cite{Zhang12p1425} of the stability of ABC
compounds, sixteen new combinations are predicted to form in the
half-Heusler structure. Of these, fifteen are in our search set, and
we find that six are insulating. Our results for these compounds are
given in Table \ref{table:Table3}, where the compounds are ranked by
$k_{14}$. It should be noted, however, that 462 of the 987
combinations in our search set were not considered in the stability
study; if they had been, the number of compounds in Table
\ref{table:Table3} would likely be larger and span a wider range of
property values.  In any case, absence from this list does not preclude
the synthesis of the half-Heusler structure as a metastable phase with
appropriate processing.
For example, our calculations on KBaN indicate that the half-Heusler
structure is the second-lowest in energy among seven structures
studied, and is only 0.11\,eV higher than the lowest-energy ($Pnma$)
structure; such a case could be promising candidate for directed
synthesis as a metastable phase.

In summary, we have used a first-principles rational-design approach
to demonstrate semiconducting half-Heusler compounds as a previously
unrecognized class of piezoelectric materials. We have presented these
predictions first for compounds already reported in the half-Heusler
structure, and then for a much larger set of candidate combinations
that were generated and screened for high performance via a
high-throughput analysis.  We hope that our results will provide
guidance for the experimental realization and further investigation of
high-performance materials suitable for practical applications.  We
also suggest that the combination of piezoelectric properties with
other characteristic properties of Heuslers, especially magnetic
properties, may offer inviting avenues for further development of
multifunctional materials.

\vspace{0.3cm}
\noindent{\bf Acknowledgments}
\vspace{0.3cm}

This work was supported in part by ONR Grant N00014-05-1-0054 and
N000-14-09-1-0300.  Calculations were carried out at the Center for
Piezoelectrics by Design.  We thank S.-W. Cheong, C. Felser,
K. Garrity, D. R. Hamann, K. Haule, I. Takeuchi, Zhiping Yin and Qibin
Zhou for useful discussions.


\begin{thebibliography}{28}
\expandafter\ifx\csname natexlab\endcsname\relax\def\natexlab#1{#1}\fi
\expandafter\ifx\csname bibnamefont\endcsname\relax
  \def\bibnamefont#1{#1}\fi
\expandafter\ifx\csname bibfnamefont\endcsname\relax
  \def\bibfnamefont#1{#1}\fi
\expandafter\ifx\csname citenamefont\endcsname\relax
  \def\citenamefont#1{#1}\fi
\expandafter\ifx\csname url\endcsname\relax
  \def\url#1{\texttt{#1}}\fi
\expandafter\ifx\csname urlprefix\endcsname\relax\def\urlprefix{URL }\fi
\providecommand{\bibinfo}[2]{#2}
\providecommand{\eprint}[2][]{\url{#2}}

\bibitem[{\citenamefont{Wu et~al.}(2005)\citenamefont{Wu, Vanderbilt, and
  Hamann}}]{Wu05p035105}
\bibinfo{author}{\bibfnamefont{X.}~\bibnamefont{Wu}},
  \bibinfo{author}{\bibfnamefont{D.}~\bibnamefont{Vanderbilt}},
  \bibnamefont{and} \bibinfo{author}{\bibfnamefont{D.~R.}
  \bibnamefont{Hamann}}, \bibinfo{journal}{Phys. Rev. B}
  \textbf{\bibinfo{volume}{72}}, \bibinfo{pages}{035105}
  (\bibinfo{year}{2005}).

\bibitem[{\citenamefont{Baroni et~al.}(2001)\citenamefont{Baroni,
  de{}Gironcoli, Dal{}Corso, and Giannozzi}}]{Baroni01p515}
\bibinfo{author}{\bibfnamefont{S.}~\bibnamefont{Baroni}},
  \bibinfo{author}{\bibfnamefont{S.}~\bibnamefont{de{}Gironcoli}},
  \bibinfo{author}{\bibfnamefont{A.}~\bibnamefont{Dal{}Corso}},
  \bibnamefont{and}
  \bibinfo{author}{\bibfnamefont{P.}~\bibnamefont{Giannozzi}},
  \bibinfo{journal}{Rev. Mod. Phys.} \textbf{\bibinfo{volume}{73}},
  \bibinfo{pages}{515} (\bibinfo{year}{2001}).

\bibitem[{\citenamefont{Fischer et~al.}(2006)\citenamefont{Fischer, Tibbetts,
  Morgan, and Ceder}}]{Fischer06p641}
\bibinfo{author}{\bibfnamefont{C.~C.} \bibnamefont{Fischer}},
  \bibinfo{author}{\bibfnamefont{K.~J.} \bibnamefont{Tibbetts}},
  \bibinfo{author}{\bibfnamefont{D.}~\bibnamefont{Morgan}}, \bibnamefont{and}
  \bibinfo{author}{\bibfnamefont{G.}~\bibnamefont{Ceder}},
  \bibinfo{journal}{Nature Mater.} \textbf{\bibinfo{volume}{5}},
  \bibinfo{pages}{641} (\bibinfo{year}{2006}).

\bibitem[{\citenamefont{Hautier et~al.}(2010)\citenamefont{Hautier, Fischer,
  Jain, Mueller, and Ceder}}]{Hautier10p3762}
\bibinfo{author}{\bibfnamefont{G.}~\bibnamefont{Hautier}},
  \bibinfo{author}{\bibfnamefont{C.~C.} \bibnamefont{Fischer}},
  \bibinfo{author}{\bibfnamefont{A.}~\bibnamefont{Jain}},
  \bibinfo{author}{\bibfnamefont{T.}~\bibnamefont{Mueller}}, \bibnamefont{and}
  \bibinfo{author}{\bibfnamefont{G.}~\bibnamefont{Ceder}},
  \bibinfo{journal}{Chem. Mater.} \textbf{\bibinfo{volume}{22}},
  \bibinfo{pages}{3762} (\bibinfo{year}{2010}).

\bibitem[{\citenamefont{Nowotny and Bachmayer}(1950)}]{Nowotny50p488}
\bibinfo{author}{\bibfnamefont{H.}~\bibnamefont{Nowotny}} \bibnamefont{and}
  \bibinfo{author}{\bibfnamefont{K.}~\bibnamefont{Bachmayer}},
  \bibinfo{journal}{Monatsch. Chem.} \textbf{\bibinfo{volume}{81}},
  \bibinfo{pages}{488} (\bibinfo{year}{1950}).

\bibitem[{\citenamefont{Belsky et~al.}(2002)\citenamefont{Belsky, Hellenbrandt,
  Karen, and Luksch}}]{Belsky02p364}
\bibinfo{author}{\bibfnamefont{A.}~\bibnamefont{Belsky}},
  \bibinfo{author}{\bibfnamefont{M.}~\bibnamefont{Hellenbrandt}},
  \bibinfo{author}{\bibfnamefont{V.~L.} \bibnamefont{Karen}}, \bibnamefont{and}
  \bibinfo{author}{\bibfnamefont{P.}~\bibnamefont{Luksch}},
  \bibinfo{journal}{Acta Cryst.} \textbf{\bibinfo{volume}{B58}},
  \bibinfo{pages}{364} (\bibinfo{year}{2002}).

\bibitem[{\citenamefont{Kandpal et~al.}(2006)\citenamefont{Kandpal, Felser, and
  Seshadri}}]{Kandpal06p776}
\bibinfo{author}{\bibfnamefont{H.~C.} \bibnamefont{Kandpal}},
  \bibinfo{author}{\bibfnamefont{C.}~\bibnamefont{Felser}}, \bibnamefont{and}
  \bibinfo{author}{\bibfnamefont{R.}~\bibnamefont{Seshadri}},
  \bibinfo{journal}{J. Phys. D.} \textbf{\bibinfo{volume}{39}},
  \bibinfo{pages}{776} (\bibinfo{year}{2006}).

\bibitem[{\citenamefont{de{}Groot et~al.}(1983)\citenamefont{de{}Groot,
  Mueller, van{}Engen, and Buschow}}]{deGroot83p2024}
\bibinfo{author}{\bibfnamefont{R.~A.} \bibnamefont{de{}Groot}},
  \bibinfo{author}{\bibfnamefont{F.~M.} \bibnamefont{Mueller}},
  \bibinfo{author}{\bibfnamefont{P.~G.} \bibnamefont{van{}Engen}},
  \bibnamefont{and} \bibinfo{author}{\bibfnamefont{K.~H.~J.}
  \bibnamefont{Buschow}}, \bibinfo{journal}{Phys. Rev. Lett.}
  \textbf{\bibinfo{volume}{50}}, \bibinfo{pages}{2024} (\bibinfo{year}{1983}).

\bibitem[{\citenamefont{Chadov et~al.}(2010)\citenamefont{Chadov, Qi, Kubler,
  Fecher, Felser, and Zhang}}]{Chadov10p541}
\bibinfo{author}{\bibfnamefont{S.}~\bibnamefont{Chadov}},
  \bibinfo{author}{\bibfnamefont{X.}~\bibnamefont{Qi}},
  \bibinfo{author}{\bibfnamefont{J.}~\bibnamefont{Kubler}},
  \bibinfo{author}{\bibfnamefont{G.~H.} \bibnamefont{Fecher}},
  \bibinfo{author}{\bibfnamefont{C.}~\bibnamefont{Felser}}, \bibnamefont{and}
  \bibinfo{author}{\bibfnamefont{S.~C.} \bibnamefont{Zhang}},
  \bibinfo{journal}{Nature Mater.} \textbf{\bibinfo{volume}{9}},
  \bibinfo{pages}{541} (\bibinfo{year}{2010}).

\bibitem[{\citenamefont{Lin et~al.}(2010)\citenamefont{Lin, Wray, Xia, Xu, Jia,
  Cava, Bansil, and Hasan}}]{Lin10p546}
\bibinfo{author}{\bibfnamefont{H.}~\bibnamefont{Lin}},
  \bibinfo{author}{\bibfnamefont{A.}~\bibnamefont{Wray}},
  \bibinfo{author}{\bibfnamefont{Y.}~\bibnamefont{Xia}},
  \bibinfo{author}{\bibfnamefont{S.}~\bibnamefont{Xu}},
  \bibinfo{author}{\bibfnamefont{S.}~\bibnamefont{Jia}},
  \bibinfo{author}{\bibfnamefont{R.~J.} \bibnamefont{Cava}},
  \bibinfo{author}{\bibfnamefont{A.}~\bibnamefont{Bansil}}, \bibnamefont{and}
  \bibinfo{author}{\bibfnamefont{M.~Z.} \bibnamefont{Hasan}},
  \bibinfo{journal}{Nature Mater.} \textbf{\bibinfo{volume}{9}},
  \bibinfo{pages}{546} (\bibinfo{year}{2010}).

\bibitem[{\citenamefont{Felser et~al.}(2007)\citenamefont{Felser, Fecher, and
  Balke}}]{Felser07p668}
\bibinfo{author}{\bibfnamefont{C.}~\bibnamefont{Felser}},
  \bibinfo{author}{\bibfnamefont{G.~H.} \bibnamefont{Fecher}},
  \bibnamefont{and} \bibinfo{author}{\bibfnamefont{B.}~\bibnamefont{Balke}},
  \bibinfo{journal}{Angew. Chem. Int. Ed.} \textbf{\bibinfo{volume}{46}},
  \bibinfo{pages}{668} (\bibinfo{year}{2007}).

\bibitem[{\citenamefont{Shen et~al.}(2001)\citenamefont{Shen, Chen, Goto,
  Hirai, Yang, Meisner, and Uher}}]{Shen01p4165}
\bibinfo{author}{\bibfnamefont{Q.}~\bibnamefont{Shen}},
  \bibinfo{author}{\bibfnamefont{L.}~\bibnamefont{Chen}},
  \bibinfo{author}{\bibfnamefont{T.}~\bibnamefont{Goto}},
  \bibinfo{author}{\bibfnamefont{T.}~\bibnamefont{Hirai}},
  \bibinfo{author}{\bibfnamefont{J.}~\bibnamefont{Yang}},
  \bibinfo{author}{\bibfnamefont{G.~P.} \bibnamefont{Meisner}},
  \bibnamefont{and} \bibinfo{author}{\bibfnamefont{C.}~\bibnamefont{Uher}},
  \bibinfo{journal}{Appl. Phys. Lett.} \textbf{\bibinfo{volume}{79}},
  \bibinfo{pages}{4165} (\bibinfo{year}{2001}).

\bibitem[{\citenamefont{Nolas et~al.}(2006)\citenamefont{Nolas, Poon, and
  Kanatzidis}}]{Nolas06p199}
\bibinfo{author}{\bibfnamefont{G.~S.} \bibnamefont{Nolas}},
  \bibinfo{author}{\bibfnamefont{J.}~\bibnamefont{Poon}}, \bibnamefont{and}
  \bibinfo{author}{\bibfnamefont{M.~G.} \bibnamefont{Kanatzidis}},
  \bibinfo{journal}{MRS Bull.} \textbf{\bibinfo{volume}{31}},
  \bibinfo{pages}{199} (\bibinfo{year}{2006}).

\bibitem[{\citenamefont{Balke et~al.}(2011)\citenamefont{Balke, Barth, Schwall,
  Fecher, and Felser}}]{Balke11p702}
\bibinfo{author}{\bibfnamefont{B.}~\bibnamefont{Balke}},
  \bibinfo{author}{\bibfnamefont{J.}~\bibnamefont{Barth}},
  \bibinfo{author}{\bibfnamefont{M.}~\bibnamefont{Schwall}},
  \bibinfo{author}{\bibfnamefont{G.~H.} \bibnamefont{Fecher}},
  \bibnamefont{and} \bibinfo{author}{\bibfnamefont{C.}~\bibnamefont{Felser}},
  \bibinfo{journal}{J. Elec. Mat.} \textbf{\bibinfo{volume}{40}},
  \bibinfo{pages}{702} (\bibinfo{year}{2011}).

\bibitem[{\citenamefont{Ogut and Rabe}(1995)}]{Ogut95p10443}
\bibinfo{author}{\bibfnamefont{S.}~\bibnamefont{Ogut}} \bibnamefont{and}
  \bibinfo{author}{\bibfnamefont{K.~M.} \bibnamefont{Rabe}},
  \bibinfo{journal}{Phys. Rev. B.} \textbf{\bibinfo{volume}{51}},
  \bibinfo{pages}{10443} (\bibinfo{year}{1995}).

\bibitem[{\citenamefont{Wood et~al.}(1985)\citenamefont{Wood, Zunger, and
  de{}Groot}}]{Wood85p2570}
\bibinfo{author}{\bibfnamefont{D.~M.} \bibnamefont{Wood}},
  \bibinfo{author}{\bibfnamefont{A.}~\bibnamefont{Zunger}}, \bibnamefont{and}
  \bibinfo{author}{\bibfnamefont{R.}~\bibnamefont{de{}Groot}},
  \bibinfo{journal}{Phys. Rev. B.} \textbf{\bibinfo{volume}{31}},
  \bibinfo{pages}{2570} (\bibinfo{year}{1985}).

\bibitem[{\citenamefont{Gonze et~al.}(2002)\citenamefont{Gonze, Beuken,
  Caracas, Detraux, Fuchs, Rignanese, Sindic, Verstraete, Zerah, Jollet
  et~al.}}]{Gonze02p478}
\bibinfo{author}{\bibfnamefont{X.}~\bibnamefont{Gonze}},
  \bibinfo{author}{\bibfnamefont{J.-M.} \bibnamefont{Beuken}},
  \bibinfo{author}{\bibfnamefont{R.}~\bibnamefont{Caracas}},
  \bibinfo{author}{\bibfnamefont{F.}~\bibnamefont{Detraux}},
  \bibinfo{author}{\bibfnamefont{M.}~\bibnamefont{Fuchs}},
  \bibinfo{author}{\bibfnamefont{G.-M.} \bibnamefont{Rignanese}},
  \bibinfo{author}{\bibfnamefont{L.}~\bibnamefont{Sindic}},
  \bibinfo{author}{\bibfnamefont{M.}~\bibnamefont{Verstraete}},
  \bibinfo{author}{\bibfnamefont{G.}~\bibnamefont{Zerah}},
  \bibinfo{author}{\bibfnamefont{F.}~\bibnamefont{Jollet}},
  \bibnamefont{et~al.}, \bibinfo{journal}{Comp. Mater. Sci.}
  \textbf{\bibinfo{volume}{25}}, \bibinfo{pages}{478} (\bibinfo{year}{2002}).

\bibitem[{\citenamefont{Gonze et~al.}(2005)\citenamefont{Gonze, Rignanese,
  Verstraete, Beuken, Pouillon, Caracas, Jollet, Torrent, Zerah, Mikami
  et~al.}}]{Gonze05p558}
\bibinfo{author}{\bibfnamefont{X.}~\bibnamefont{Gonze}},
  \bibinfo{author}{\bibfnamefont{G.}~\bibnamefont{Rignanese}},
  \bibinfo{author}{\bibfnamefont{M.}~\bibnamefont{Verstraete}},
  \bibinfo{author}{\bibfnamefont{J.}~\bibnamefont{Beuken}},
  \bibinfo{author}{\bibfnamefont{Y.}~\bibnamefont{Pouillon}},
  \bibinfo{author}{\bibfnamefont{R.}~\bibnamefont{Caracas}},
  \bibinfo{author}{\bibfnamefont{F.}~\bibnamefont{Jollet}},
  \bibinfo{author}{\bibfnamefont{M.}~\bibnamefont{Torrent}},
  \bibinfo{author}{\bibfnamefont{G.}~\bibnamefont{Zerah}},
  \bibinfo{author}{\bibfnamefont{M.}~\bibnamefont{Mikami}},
  \bibnamefont{et~al.}, \bibinfo{journal}{Z. Kristall.}
  \textbf{\bibinfo{volume}{220}}, \bibinfo{pages}{558} (\bibinfo{year}{2005}).

\bibitem[{\citenamefont{Gonze et~al.}(2009)\citenamefont{Gonze, Amadon,
  Anglade, Beuken, Bottin, Boulanger, Bruneval, Caliste, Caracas, Cote
  et~al.}}]{Gonze09p2582}
\bibinfo{author}{\bibfnamefont{X.}~\bibnamefont{Gonze}},
  \bibinfo{author}{\bibfnamefont{B.}~\bibnamefont{Amadon}},
  \bibinfo{author}{\bibfnamefont{P.}~\bibnamefont{Anglade}},
  \bibinfo{author}{\bibfnamefont{J.~M.} \bibnamefont{Beuken}},
  \bibinfo{author}{\bibfnamefont{F.}~\bibnamefont{Bottin}},
  \bibinfo{author}{\bibfnamefont{P.}~\bibnamefont{Boulanger}},
  \bibinfo{author}{\bibfnamefont{F.}~\bibnamefont{Bruneval}},
  \bibinfo{author}{\bibfnamefont{D.}~\bibnamefont{Caliste}},
  \bibinfo{author}{\bibfnamefont{R.}~\bibnamefont{Caracas}},
  \bibinfo{author}{\bibfnamefont{M.}~\bibnamefont{Cote}}, \bibnamefont{et~al.},
  \bibinfo{journal}{Comp. Phys. Comm.} \textbf{\bibinfo{volume}{180}},
  \bibinfo{pages}{2582} (\bibinfo{year}{2009}).

\bibitem[{\citenamefont{Monkhorst and Pack}(1976)}]{Monkhorst76p5188}
\bibinfo{author}{\bibfnamefont{H.~J.} \bibnamefont{Monkhorst}}
  \bibnamefont{and} \bibinfo{author}{\bibfnamefont{J.~D.} \bibnamefont{Pack}},
  \bibinfo{journal}{Phys. Rev. B} \textbf{\bibinfo{volume}{13}},
  \bibinfo{pages}{5188} (\bibinfo{year}{1976}).

\bibitem[{\citenamefont{Rappe et~al.}(1990)\citenamefont{Rappe, Rabe, Kaxiras,
  and Joannopoulos}}]{Rappe90p1227}
\bibinfo{author}{\bibfnamefont{A.~M.} \bibnamefont{Rappe}},
  \bibinfo{author}{\bibfnamefont{K.~M.} \bibnamefont{Rabe}},
  \bibinfo{author}{\bibfnamefont{E.}~\bibnamefont{Kaxiras}}, \bibnamefont{and}
  \bibinfo{author}{\bibfnamefont{J.~D.} \bibnamefont{Joannopoulos}},
  \bibinfo{journal}{Phys. Rev. B Rapid Comm.} \textbf{\bibinfo{volume}{41}},
  \bibinfo{pages}{1227} (\bibinfo{year}{1990}).

\bibitem[{\citenamefont{Ramer and Rappe}(1999)}]{Ramer99p12471}
\bibinfo{author}{\bibfnamefont{N.~J.} \bibnamefont{Ramer}} \bibnamefont{and}
  \bibinfo{author}{\bibfnamefont{A.~M.} \bibnamefont{Rappe}},
  \bibinfo{journal}{Phys. Rev. B} \textbf{\bibinfo{volume}{59}},
  \bibinfo{pages}{12471} (\bibinfo{year}{1999}).

\bibitem[{opi()}]{opium}
\bibinfo{howpublished}{http://opium.sourceforge.net}.

\bibitem[{\citenamefont{Gonze and Lee}(1997)}]{Gonze97p10355}
\bibinfo{author}{\bibfnamefont{X.}~\bibnamefont{Gonze}} \bibnamefont{and}
  \bibinfo{author}{\bibfnamefont{C.}~\bibnamefont{Lee}},
  \bibinfo{journal}{Phys. Rev. B} \textbf{\bibinfo{volume}{55}},
  \bibinfo{pages}{10355} (\bibinfo{year}{1997}).

\bibitem[{\citenamefont{Born and Huang}(1954)}]{born54p268}
\bibinfo{author}{\bibfnamefont{M.}~\bibnamefont{Born}} \bibnamefont{and}
  \bibinfo{author}{\bibfnamefont{K.}~\bibnamefont{Huang}},
  \emph{\bibinfo{title}{Dynamical Theory of Crystal Lattices}}
  (\bibinfo{publisher}{Oxford University Press, Oxford}, \bibinfo{year}{1954}),
  \bibinfo{note}{we used Eq. (33.28) from this book.}

\bibitem[{\citenamefont{Hamann et~al.}(2005)\citenamefont{Hamann, Wu, Rabe, and
  Vanderbilt}}]{Hamann05p035117}
\bibinfo{author}{\bibfnamefont{D.~R.} \bibnamefont{Hamann}},
  \bibinfo{author}{\bibfnamefont{X.}~\bibnamefont{Wu}},
  \bibinfo{author}{\bibfnamefont{K.~M.} \bibnamefont{Rabe}}, \bibnamefont{and}
  \bibinfo{author}{\bibfnamefont{D.}~\bibnamefont{Vanderbilt}},
  \bibinfo{journal}{Phys. Rev. B} \textbf{\bibinfo{volume}{71}},
  \bibinfo{pages}{035117} (\bibinfo{year}{2005}).

\bibitem[{\citenamefont{Karki et~al.}(1997)\citenamefont{Karki, Ackland, and
  Crain}}]{Karki97p8579}
\bibinfo{author}{\bibfnamefont{B.~B.} \bibnamefont{Karki}},
  \bibinfo{author}{\bibfnamefont{G.~J.} \bibnamefont{Ackland}},
  \bibnamefont{and} \bibinfo{author}{\bibfnamefont{J.}~\bibnamefont{Crain}},
  \bibinfo{journal}{Journal of Physics: Condensed Matter}
  \textbf{\bibinfo{volume}{9}}, \bibinfo{pages}{8579} (\bibinfo{year}{1997}).

\bibitem[{\citenamefont{Zhang et~al.}(2012)\citenamefont{Zhang, Yu, Zakutayev,
  and Zunger}}]{Zhang12p1425}
\bibinfo{author}{\bibfnamefont{X.}~\bibnamefont{Zhang}},
  \bibinfo{author}{\bibfnamefont{L.}~\bibnamefont{Yu}},
  \bibinfo{author}{\bibfnamefont{A.}~\bibnamefont{Zakutayev}},
  \bibnamefont{and} \bibinfo{author}{\bibfnamefont{A.}~\bibnamefont{Zunger}},
  \bibinfo{journal}{Adv. Funct. Mater.} \textbf{\bibinfo{volume}{22}},
  \bibinfo{pages}{1425} (\bibinfo{year}{2012}).

\end{thebibliography}

\end{document}